\documentclass[aps, prd, superscriptaddress, nofootinbib, preprintnumbers, showpacs, showkeys]{revtex4}
\usepackage{graphicx}
\usepackage{amsmath}
\usepackage[usenames]{color}
\usepackage{allpurpose}

\begin{document}

\title{Summing Radiative Corrections to the Effective Potential}

\author{F.A. Chishtie}
\affiliation{Department of Applied
Mathematics, University of Western Ontario, London, Ontario N6A 5B7,
Canada}

\author{T. Hanif}
\affiliation{Department of Physics and Astronomy, University of Western Ontario, London, Ontario N6A 3K7,
Canada}

\author{Junji Jia}
\email{jjia5@uwo.ca}
\affiliation{Department of Applied
Mathematics, University of Western Ontario, London, Ontario N6A 5B7,
Canada}

\author{D.G.C. McKeon}
\affiliation{Department of Applied
Mathematics, University of Western Ontario, London, Ontario N6A 5B7,
Canada}
\affiliation{Department of Mathematics and Computer Science, Algoma University, Sault Ste Marie, Ontario, P6A 2G4,
Canada}

\author{T.N. Sherry}
\affiliation{School of Mathematics, Statistics and Applied Mathematics,
National University of Ireland Galway, University Road, Galway Ireland}
\affiliation{School of Theoretical Physics, Dublin Institute for Advanced Study, Burlington Rd., Dublin, 4, Ireland}

\date{\today}
\vspace{2cm}

\begin{abstract}

When one uses the Coleman-Weinberg renormalization condition, the effective potential $V$ in the massless $\phi_4^4$ theory with O(N) symmetry is completely determined by the renormalization group functions. It has been shown how the $(p+1)$ order renormalization group function determine the sum of all the N$^{\mbox{\scriptsize p}}$LL order contribution to $V$
to all orders in the loop expansion. We discuss here how, in addition to fixing the N$^{\mbox{\scriptsize p}}$LL contribution to $V$, the $(p+1)$ order renormalization group functions also can be used to determine portions of the N$^{\mbox{\scriptsize p+n}}$LL contributions to $V$. When these contributions are summed to all orders, the singularity structure of \mcv is altered. An alternate
rearrangement of the contributions to $V$ in powers of $\ln \phi$, when the extremum condition $V^\prime (\phi = v) = 0$ is combined with the renormalization group equation, show that either
$v = 0$ or $V$ is independent of $\phi$.  This conclusion is supported by showing the LL, $\cdots$, N$^4$LL contributions to $V$ become progressively less dependent on $\phi$.

\end{abstract}



\maketitle

\section{Introduction}

It is well known that the renormalization group (RG) equation fixes the relationship between different contributions to the effective potential \cite{c1,c2,c3,c4} when using perturbation theory.  In the
massless O(N) symmetric $\lambda\phi^4$ model when one uses the Coleman-Weinberg renormalization scheme, the perturbative expansion of $V$ involves powers of $\ln \frac{\phi^2}{\mu^2}$ where
$\mu$ is the renormalization scale; the $n$-loop contribution to $V$ is proportional to $\lambda^{n+1}$ and powers of $\ln \frac{\phi^2}{\mu^2}$ up to and including
$\ln^n \frac{\phi^2}{\mu^2}$.  The n-loop RG functions for this
model ($\beta(\lambda)$, the usual beta function, and $\gamma(\lambda)$, the anomalous dimension) are known to be related to the $n$-loop expression for $V$; it has also been shown that
the $n$-loop expressions for $\beta(\lambda)$ and $\gamma(\lambda)$ are completely determined by the $(n+1)$-loop expression for $V$, and furthermore, once $\beta(\lambda)$ and
$\gamma(\lambda)$ are known, portions of $V$ beyond order $(n+1)$ are fixed \cite{c5}.  If one were to use the Coleman-Weinberg (CW) renormalization scheme \cite{c1}, then these portions of $V$ can
be summed systematically \cite{c6} without the appearance of any unknown parameter.  If the one-loop RG functions are known, then the contributions to $V$ containing the highest power of
$\ln \frac{\phi^2}{\mu^2}$ at each order of perturbation theory can be summed exactly.  This is known as the ``leading-log'' (LL) sum.  In general, if the RG functions are known at $n$-loop
order in the CW renormalization scheme, then the contributions to the m$^{th}$ loop contribution to $V$ containing $(m-n+1)$ powers of $\ln \frac{\phi^2}{\mu^2}$ can be summed -- this is
the N$^{n-1}$LL sum.  Consequently, if the RG functions were known exactly in the CW scheme, then $V$ would be completely fixed.  (Here, NLL denotes the ``next-to-leading-log'' contributions
to $V$, that is, those contributions to $V$ coming at each order in the loop expansion that are next to the leading order in powers of $\ln \frac{\phi^2}{\mu^2}$; similarly N$^2$LL, N$^3$LL
contributions to $V$ are defined.)  A similar result holds for the kinetic term for the effective action in these models \cite{c7}.

However, if one were to consider the RG equation with RG functions at $n$-loop order, the solution to this equation is not given by the sum of the LL, NLL, $\cdots$, N$^{\mbox{\scriptsize n-1}}$LL contributions
to $V$, even though these contributions are determined by this equation.  The exact solution to this RG equation with the RG functions given to $n$-loop order in fact can also fix portions
of the N$^{\mbox{\scriptsize n}}$LL, N$^{\mbox{\scriptsize n+1}}$LL, $\cdots$ contributions to $V$.  In this paper, we systematically examine how portions of these higher order contributions to $V$ can be determined and then summed.
These sums reveal a singularity structure in $V$ that is different from the usual ``Landau singularity'' apparent in the individual LL, NLL, $\cdots$ contributions to $V$.

We next consider another rearrangement of the perturbative contributions to $V$, this time expanding $V$ in powers of $\ln \frac{\phi^2}{\mu^2}$ with the coefficients being dependent
solely on the coupling $\lambda$ with contributions coming from all orders of the loop expansion.  The RG equation fixes the coefficient of the n$^{th}$ power of $\ln \frac{\phi^2}{\mu^2}$
in terms of the coefficient of the $(n-1)^{st}$ power of $\ln \frac{\phi^2}{\mu^2}$; by iterating this dependency, all of these coefficients can be expressed in terms of the
contribution to $V$ independent of $\ln \frac{\phi^2}{\mu^2}$. A second condition, that $V$ have an extremum when $\phi = v$, serves to fix this $\log$-independent piece of $V$.
Remarkably, this has the consequence that $V$ becomes independent of $\phi$, unless $v = 0$, in which case there is no spontaneous symmetry breaking.  This argument has been presented in
detail in Ref. \cite{c8} in the context of using modified minimal subtraction ($\overline{\mbox{MS}}$) to renormalize $V$, and not only for massless $\lambda\phi^4$ theory; scalar quantum electrodynamics and massive
$\lambda\phi^4$ theory were also found to have this property.  In this paper we reexamine the massless $\lambda\phi^4$ model, this time using the CW renormalization scheme.  Here we find
that not only is the potential ``flat'', but also that the coupling vanishes.  The model is consequently ``trivial''; this feature has previously been
discussed in Ref. \cite{c15}.  We then look at the full LL, NLL, $\cdots$, N$^4$LL contributions to $V$ in the massless $\lambda\phi^4$ theory as
considered in Ref. \cite{c10} and show graphically that as $p$ increases from zero to four, the N$^{\mbox{\scriptsize p}}$LL contributions to $V$ give a progressively better approximation to $V$ being independent of
$\phi$.  We also consider the N$^{\mbox{\scriptsize p}}$LL (p = 0,1,2) contributions to $V$ in the $\overline{\mbox{MS}}$ scheme and find that this pattern reoccurs.

We note that the potential we are considering is the sum of all one particle irreducible diagrams with no external momentum.  This is not the ``effective potential'', which is
convex and real, that has been discussed in Ref. \cite{c17} and reviewed in \cite{c18,c19}.  The potential we are discussing here is the one relevant for analyzing spontaneous symmetry breakdown.

\section{Finding N$^{\mbox{\scriptsize p}}$LL Contribution to $V$}
In an O(N)-symmetric scalar model with a massless potential
\be V_{\mbox{\scriptsize cl}}=\lambda\phi^4~,\ee
radiative corrections to the potential are of the form \cite{c1,c2,c3,c4}
\be V=\sum_{n=0}^\infty\sum_{m=0}^n\lambda^{n+1}T_{n,m}L^m\phi^4 \label{e2}\ee
where $L=\ln(\phi^2/\mu^2)$ and the CW RG condition \cite{c1}
\be \left. \frac{\dd^4 V}{\dd\phi^4}\right|_{\phi=\mu}=24\lambda\label{e3}\ee
has been used. The $n$-loop contribution to $V$ fix the coefficients $T_{n,m}~(m \leq n)$.
\mcv is independent of the unphysical renormalization scale parameter \mmu provided the RG equation is satisfied
\be \lb \mu\frac{\partial}{\partial \mu}+\beta(\lambda)\frac{\partial}{\partial \lambda}+\gamma(\lambda)\phi\frac{\partial}{\partial \phi}\rb V(\lambda,\phi,\mu)=0. \label{e4}\ee
The RG functions $\displaystyle \beta(\lambda)=\mu\frac{\dd \lambda}{\dd\mu}$, $\displaystyle \gamma(\lambda)=\frac{\mu}{\phi}\frac{\dd \phi}{\dd\mu}$ are needed to find the implicit dependence of \mcv on $\mu$. Upon expanding
\bea \beta(\lambda)&=&\sum_{k=2}^\infty b_k\lambda^k, \\
\gamma(\lambda)&=&\sum_{k=1}^\infty g_k\lambda^k \eea
(where $b_{k-1}$ and $g_k$ come from k-loop considerations) and regrouping the sum in eq. \refer{e2} so that\\ $V = \displaystyle{\sum_{n=0}^\infty} \lambda^{n+1} S_n(\xi)\phi^4$ with
\be S_n(\xi)=\sum_{m=0}^\infty T_{n+m,m}\xi^m, \qquad \xi=\lambda L\label{e6}\ee
then eq. \refer{e4} is satisfied at order $\lambda^{n+2}$ if
\begin{subequations}\label{e7}
\bea && \lsb (-2+b_2\xi)\frac{\dd}{\dd\xi}+(b_2+4g_1)\rsb S_0(\xi)=0, \\
    \mbox{and } && \lsb (-2+b_2\xi)\frac{\dd}{\dd \xi}+(n+1)b_2+4g_1\rsb S_n(\xi)\nonumber\\
     &&+\sum_{m=0}^{n-1}\lsb (2g_{n-m}+b_{n-m+2}\xi)\frac{\dd}{\dd \xi}+(m+1)b_{n+2-m}+4g_{n+1-m}\rsb S_m(\xi)=0. \eea
\end{subequations}
These nested equations can be solved in turn for $S_0, ~S_1,~S_2,~\cdots,$ with the boundary conditions $S_n(0)=T_{n,0}$; in particular
\bea S_0&=&\frac{T_{0,0}}{w},\label{eq9} \\
S_1&=&-\frac{4g_2T_{0,0}}{b_2w}+\frac{4g_2T_{0,0}+b_2T_{1,0}}{b_2w^2}-\frac{b_3T_{0,0}}{b_2w^2}\ln|w| \eea where
\be w=1-\frac{b_2}{2}\xi. \label{eq10}\ee
The sum for $S_n(\xi)$ gives the total N$^n$LL contribution to $V$; it contains portions of the $p$-loop contribution to $V$ for all $p$.
Eq. \refer{e3} can be used to determine the boundary values $T_{n,0}$; we find from eqs. \refer{e3} and \refer{e6} that $S_0(0)=T_{0,0}=1$ and
\be 16S_k''''(0)+80S_{k+1}'''(0)+140S_{k+2}''(0)+100S_{k+3}'(0)+24S_{k+4}(0)=0\label{e10}\ee
for $k=0,~1,~\cdots$ from which we can find $T_{k+1,0}(0)$ once $S_k(0),~\cdots,~S_0(0)$ are known.

The RG function in the minimal subtraction (MS) scheme, $\tilde{\beta}(\lambda),~\tilde{\gamma}(\lambda)$, have been computed to five-loop order \cite{c9}. In this scheme, the expansion of \refer{e2} for \mcv involves the logarithm $\tilde{L}=\ln(\lambda\phi^2/\tilde{\mu}^2)$ where $\tilde{\mu}$ is the scale parameter in the MS scheme. By the rescaling $\tilde{\mu}=\mu\sqrt{\lambda}$ one can go from the MS to the CW scheme, and as $\displaystyle \frac{\dd\mu}{\dd\tilde{\mu}}=\lambda^{-1/2}\lb 1-\frac{\tilde{\beta}(\lambda)}{2\lambda}\rb $ we find \cite{c11}
\be \beta(\lambda)=\frac{\tilde{\beta}(\lambda)}{\displaystyle 1-\frac{\tilde{\beta}(\lambda)}{2\lambda}},\quad \gamma(\lambda)=\frac{\tilde{\gamma}(\lambda)}{\displaystyle 1-\frac{\tilde{\beta}(\lambda)}{2\lambda}},\label{e13v2}\ee where $\displaystyle \tilde{\beta}(\lambda)=\tilde{\mu}\frac{\dd \lambda}{\dd\tilde{\mu}}$, $\displaystyle \tilde{\gamma}(\lambda)=\frac{\tilde{\mu}}{\phi}\frac{\dd \phi}{\dd\tilde{\mu}}$. These relations allow us to convert the MS RG functions to the CW RG functions.

We note that simply substituting $\tilde{\mu}=\mu\sqrt{\lambda}$ in the MS expansion of \mcv does not necessarily mean that the renormalization condition of eq. \refer{e3} (and consequently \refer{e10}) is satisfied. However, a finite renormalization of $\lambda$ (i.e., $\lambda\to\lambda(1+l_1\lambda+l_2\lambda^2+\cdots)$) and $\phi$ (i.e., $\phi\to\phi(1+f_1\lambda+f_2\lambda^2+\cdots)$) can always be used to ensure that these equations are satisfied without altering the value of $T_{n,m}$ (m>0). Consequently eq. \refer{e13v2} can be used to convert the RG functions from the MS to the CW scheme.

From eq. \refer{e7} it is apparent that the general solution for $S_n(\xi)$ is \cite{c6,c12}
\be S_n(\xi)=\frac{1}{b_2}\sum_{i=1}^{n+1}\sum_{j=0}^{i-1}\sigma^n_{i,j}\frac{\Lambda^j}{w^i}\label{e12}\ee
where $\Lambda\equiv \ln|w|$. It is this form of $S_n(\xi)$ that we now turn to. We now will show that if the RG functions are known to order $n$, then not only are $S_0,~ \cdots,~ S_{n-1}$
completely determined in the CW scheme, but also portions of $S_n,~ S_{n+1} \cdots$ and that these contributions can be summed.

\section{Contributions to $S_n(\xi)$}

If eq. \refer{e12} is substituted into eq. \refer{e7}, we find the recursion relation \cite{c6}
\bea 0&=&b_2(j+1)\sigma^n_{i,j+1}+[(n-i+1)b_2+4g_1]\sigma^n_{i,j}\nonumber\\
&&+\sum_{m=0}^{n-1}\lsb (j+1)b_{n+2-m}\sigma^m_{i,j+1}+(i-1)(b_2g_{n-m}+b_{n+2-m})\sigma^m_{i-1,j}\right.\nonumber\\
&&\left.-(j+1)(b_2g_{n-m}+b_{n+2-m})\sigma^m_{i-1,j+1}+(4g_{n+1-m}+(m-i+1)b_{n+2-m})\sigma^m_{i,j}\rsb,\label{e13}\eea
with $\sigma^n_{i,j}=0$ if $i>n+1,~j>i-1,~i<1,~j<0$.

We now set $j=i-1$ in eq. \refer{e13} so that
\begin{subequations} \label{e14}
\be [(n-i+1)b_2+4g_1]\sigma^n_{i,i-1}+\sum_{m=i-1}^{n-1}\lsb 4g_{n+1-m}+(m-i+1)b_{n+2-m}\rsb \sigma^m_{i,i-1}=0;\label{e14a}\ee
so also if $j=i-2$ then
\bea &&(n-i+1)b_2\sigma^n_{i,i-2}+(i-1)b_2\sigma^n_{i,i-1}+\sum_{m=i-2}^{n-1}\lsb (i-1)b_{n+2-m}\sigma^m_{i,i-1}\right.\nonumber\\
&&\left.+(i-1)(b_2g_{n-m}+b_{n+2-m})\sigma^m_{i-1,i-2}+(4g_{n+1-m}+(m-i+1)b_{n+2-m})\sigma^m_{i,i-2}\rsb=0.\label{e14b}\eea
\end{subequations}

In eq. \refer{e14a}, we take $i=n+1$ so that
\be 4g_1\sigma^n_{n+1,n}=0\ee
which gives the usual result
\be g_1=0.\ee
If now in eq. \refer{e14b} we set $i=n+1$ then
\be b_2\sigma^n_{n+1,n}+b_3\sigma^{n-1}_{n,n-1}=0\label{e17}\ee from which it follows that
\be \sigma^n_{n+1}=\rho^n\sigma^0_{1,0}~\mbox{ with }~\rho\equiv -b_3/b_2. \label{e18}\ee

We now let $i=n$ in eq. \refer{e14a} which leads to
\be b_2\sigma^n_{n,n-1}+4g_2\sigma^{n-1}_{n,n-1}=0;\label{e19}\ee
together eqs. \refer{e18} and \refer{e19} give
\be \sigma^n_{n,n-1}=-\frac{4g_2}{b_2}\rho^{n-1}\sigma^0_{1,0}.\label{e20} \ee

Next, if $i=n-1$ in eq. \refer{e14a} we get
\be 2b_2\sigma^n_{n-1,n-2}+\lsb 4g_3\sigma^{n-2}_{n-1,n-2}+(4g_2+b_3)\sigma^{n-1}_{n-1,n-2}\rsb=0\ee
which, with eqs. \refer{e18} and \refer{e20} gives
\be \sigma^n_{n-1,n-2}=-\frac{1}{2b_2}\lsb 4g_3+(4g_2+b_3)\lb -\frac{4g_2}{b_2}\rb \rsb \rho^{n-2}\sigma^0_{1,0}. \label{e22}\ee

When setting $i=n-2$ in eq. \refer{e14a} we find that
\be 3b_2\sigma^n_{n-2,n-3}+4g_4\sigma^{n-3}_{n-2,n-3}+(4g_3+b_4)\sigma^{n-2}_{n-2,n-3}+(4g_2+2b_3)\sigma^{n-1}_{n-2,n-3}=0, \ee
which by eqs. (\ref{e18}, ~\ref{e20},~\ref{e22}) becomes
\be \sigma^n_{n-2,n-3}=-\frac{1}{3b_2}\lsb 4g_4+(4g_3+b_4)\lb -\frac{4g_2}{b_2}\rb+\lb \frac{2g_2+b_3}{b_2}\rb \lb \frac{4g_2}{b_2}(4g_2+b_3)-4g_3\rb \rsb \rho^{n-3}\sigma^0_{1,0}. \label{e24}\ee
With $i=n-3$, then eq. \refer{e14a} leads to
\be 4b_2\sigma^n_{n-3,n-4}+4g_5\sigma^{n-4}_{n-3,n-4}+ (4g_4+b_5)\sigma^{n-3}_{n-3,n-4}+(4g_3+2b_4)\sigma^{n-2}_{n-3,n-4}+(4g_2+3b_3)\sigma^{n-1}_{n-3,n-4}=0 \ee
which, by eqs. (\ref{e18},~\ref{e20},~\ref{e24}) gives us
\bea \sigma^n_{n-3,n-4}&=&-\frac{1}{4b_2}\lcb 4g_5+(4g_4+b_5)\lb -\frac{4g_2}{b_2}\rb+(4g_3+2b_4)\lb\frac{-1}{2b_2}\rb\lsb 4g_3-\frac{4g_2}{b_2}(4g_2+b_3)\rsb\right.\nonumber\\
&&\left.- \frac{4g_2+3b_3}{3b_2} \lsb 4g_4-\frac{4g_2}{b_2}(4g_3+b_4)+\frac{2g_2+b_3}{b_2} \lb \frac{4g_2}{b_2}(4g_2+b_3)-4g_3\rb\rsb\rcb\rho^{n-4}\sigma^0_{1,0}.\label{e26}\eea
Setting $i=n-4$ in eq. \refer{e14a} would lead to an expression for $\sigma^n_{n-4,n-5}$ involving $g_6$, which is unknown.

We now set $i=n+1$ in eq. \refer{e13} which leads to
\be \sigma^n_{n+1,j+1}=\rho\lsb \frac{n}{j+1}\sigma^{n-1}_{n,j}-\sigma^{n-1}_{n,j+1}\rsb . \label{e27}\ee
If in eq. \refer{e27} we set $j=n-1$ we recover eq. \refer{e17}; with $j=n-2$ we find that
\be \sigma^n_{n+1,n-1}=\rho\lsb\frac{n}{n-1}\sigma^{n-1}_{n,n-2}-\sigma^{n-1}_{n,n-1}\rsb\ee
which, upon iterating and using eq. \refer{e18}, gives us
\be \sigma^n_{n+1,n-1}=n\rho^{n-1}\lsb \sigma^1_{2,0}-\rho\lb \frac{1}{2}+\frac{1}{3}+\cdots+\frac{1}{n}\rb \sigma^0_{1,0}\rsb .\label{e29}\ee
We now can set $i=n$ in eq. \refer{e14b} so that
\be b_2\sigma^n_{n,n-2}+(n-1)\lsb b_2\sigma^n_{n,n-1}+(b_2g_2+b_4)\sigma^{n-2}_{n-1,n-2}
+b_3(\sigma^{n-1}_{n,n-1}+\sigma^{n-1}_{n-1,n-2})\rsb+4g_2\sigma^{n-1}_{n,n-2}=0. \ee
Iterating this equation and using eqs. (\ref{e18},~\ref{e20},~\ref{e29}) one obtains
\bea \sigma^n_{n,n-2}&=&-\frac{1}{b_2}\lcb (n-1)\sigma^0_{1,0}\lsb \rho^{n-1}(b_2+b_3)+ \rho^{n-2}\lb (b_2g_2+b_4)
+b_3\lb\frac{-4g_2}{b_2}\rb\rb\rsb \right.\nonumber\\
&&\left.+4g_2\lsb (n-1)\rho^{n-2}\sigma^1_{2,0}-(n-1)\rho^{n-1}\lb\frac{1}{2}+\frac{1}{3}+\cdots+\frac{1}{n-1}\rb\sigma^0_{1,0}\rsb\rcb.\label{e31}\eea
With $i=n-1$ in eq. \refer{e14b}, the steps used to derive eq. \refer{e31} leads to
\bea \sigma^n_{n-1,n-3}&=&-\frac{1}{2b_2}\sigma^0_{1,0}\lcb (n-2)\lsb \rho^{n-2}\lb b_2\lb \frac{-1}{2b_2}\rb\lb4g_3-\frac{4g_2}{b_2}(4g_2+b_3)\rb +b_4+b_3\lb\frac{-4g_2}{b_2}\rb\rb\right.\right.\nonumber\\
&&\left.+\rho^{n-3}\lb(b_2g_3+b_5)+(b_2g_2+b_4)\lb\frac{-4g_2}{b_2}\rb+b_3\lb \frac{-1}{2b_2}\rb\lb4g_3-\frac{4g_2}{b_2}(4g_2+b_3)\rb \rb\rsb\nonumber\\
&&\left.+4g_3\rho^{n-2}+(4g_2+b_3)\lb\frac{-4g_2}{b_2}\rb \rho^{n-2}\rcb.\eea
It also follows from eq. \refer{e14b} if $i=n-2$ that
\bea \sigma^n_{n-2,n-4}&=&\frac{-1}{3b_2}\lcb -\frac{(n-3)}{3}\lsb4g_4+(4g_3+b_4)\lb\frac{-4g_2}{b_2}\rb +\lb\frac{2g_2+b_3}{b_2}\rb \lb\frac{4g_2}{b_2}(4g_2+b_3)-4g_3\rb\rsb \rho^{n-3}\sigma^0_{1,0}\right.\nonumber\\
&&+(n-3)b_5\rho^{n-3}\sigma^0_{1,0}+(n-3)b_4\lb\frac{-4g_2}{b_2}\rb\rho^{n-3}\sigma^0_{1,0}+(n-3)(b_2g_3+b_5)\lb\frac{-4g_2}{b_2}\rb\rho^{n-4}\sigma^0_{1,0}\nonumber\\
&&+(n-3)b_3\lb\frac{-1}{2b_2}\rb\lsb 4g_3-\frac{4g_2}{b_2}(4g_2+b_3)\rsb \rho^{n-3}\sigma^0_{1,0}+(n-3)(b_2g_4+b_6)\rho^{n-4}\sigma^0_{1,0}\nonumber\\
&&+(n-3)(b_2g_2+b_4)\lb\frac{-4g_2}{b_2}\rb\rho^{n-4}\sigma^0_{1,0}\nonumber\\
&&+(n-3)b_3\lb \frac{-1}{3b_2}\rb \lsb4g_4-\frac{4g_2}{b_2}(4g_3+b_4)+ \lb\frac{2g_2+b_3}{b_2}\rb\lb\frac{4g_2}{b_2}(4g_2+b_3)-4g_3\rb\rsb \rho^{n-4}\sigma^0_{1,0}\nonumber\\
&&+4g_4\lsb(n-3)\rho^{n-4}\sigma^1_{2,0}-(n-3)\lb\frac{1}{2}+\frac{1}{3}+\cdots+\frac{1}{n-3}\rb\rho^{n-3}\sigma^0_{1,0}\rsb\nonumber\\
&&+(4g_3+b_4)\lb\frac{-1}{b_2}\rb\lsb(n-3)\lb b_2+b_2g_2+b_4+b_3-4g_2\lb \frac{1}{2}+\frac{1}{3}+\cdots+\frac{1}{n-3}\rb\rb\rho^{n-3}\sigma^0_{1,0}\right.\nonumber\\
&&\left.+(n-3)\rho^{n-4}\lb b_3\lb\frac{-4g_2}{b_2}\rb  \sigma^0_{1,0} +4g_2\sigma^1_{2,0}\rb\rsb\nonumber\\
&&+(4g_2+2b_3)\lb\frac{-1}{2b_2}\rb\lsb (n-3) \sigma^0_{1,0} \rho^{n-3}\lb\lb\frac{2g_2}{b_2}(4g_2+b_3)-2g_3\rb +b_4+b_3\lb\frac{-4g_2}{b_2}\rb\rb\right.\nonumber\\
&&+(n-3)\sigma^0_{1,0}\rho^{n-4}\lb (b_2g_3+b_5)+(b_2g_2+b_4)\lb\frac{-4g_2}{b_2}\rb +b_3\lb\frac{-1}{2b_2}\rb\lb 4g_3-\frac{4g_2}{b_2}(4g_2+b_3)\rb\rb\nonumber\\
&&\left.\left.+\sigma^0_{1,0}\rho^{n-3}\lb4g_3+(4g_2+b_3)\lb\frac{-4g_2}{b_2}\rb\rb\rsb\rcb
\label{e33}.\eea
If one were to set $i=n-3$ in eq. \refer{e14b}, $\sigma^n_{n-3,n-5}$ would be obtained but this expression would involve $b_7$ which has not as yet been computed.

One could also set $j=n-3$ in eq. \refer{e27} so that
\be \sigma^n_{n+1,n-2}=\rho\lb\frac{n}{n-2}\sigma^{n-1}_{n,n-3}-\sigma^{n-1}_{n,n-2}\rb;\ee
iteration of this equation and using eq. \refer{e29} results in
\bea \sigma^n_{n+1,n-2}&=&n(n-1)\lcb \frac{\rho^{n-2}}{2\cdot 1}\sigma^2_{3,0}-\frac{\rho^{n-1}}{2\cdot 1}\sigma^1_{2,0}-\frac{\rho^{n-2}}{3\cdot2}\lsb -2\rho^2\lb\frac{1}{2}\rb\sigma^0_{1,0}+2\sigma^1_{2,0}\rsb\right.\nonumber\\
&&-\frac{\rho^{n-3}}{4\cdot3}\lsb-3\rho^3\lb\frac{1}{2}+\frac{1}{3}\rb\sigma^0_{1,0}+3\rho^2\sigma^1_{2,0}\rsb\nonumber\\
&&-\cdots-\frac{\rho^2}{(n-1)(n-2)}\lsb -(n-2)\rho^{n-2}\lb\frac{1}{2}+\frac{1}{3}+\cdots+\frac{1}{n-2}\rb\sigma^0_{1,0}+(n-2)\rho^{n-3}\sigma^1_{2,0}\rsb\nonumber\\
&&\left.-\frac{\rho}{n(n-1)}\lsb -(n-1)\rho^{n-1}\lb\frac{1}{2}+\frac{1}{3}+\cdots+\frac{1}{n-1}\rb\sigma^0_{1,0}+(n-1)\rho^{n-2}\sigma^1_{2,0}\rsb\rcb
\label{e35}.\eea

The contribution of $\sigma^n_{n+1,n}$ (eq. \refer{e18}), $\sigma^n_{n,n-1}$ (eq. \refer{e20}),  $\sigma^n_{n-1,n-2}$ (eq. \refer{e22}),  $\sigma^n_{n-2,n-3}$ (eq. \refer{e24}), $\sigma^n_{n-3,n-4}$ (eq. \refer{e26}),  $\sigma^n_{n+1,n-1}$ (eq. \refer{e29}),  $\sigma^n_{n,n-2}$ (eq. \refer{e31}), $\sigma^n_{n-1,n-3}$ (eq. \refer{e33})  and  $\sigma^n_{n+1,n-2}$ (eq. \refer{e35}) to \mcv can now be worked out (as could additional more complicated contributions which follow from eq. \refer{e13}). If we compute
\be V_{A,B}=\frac{1}{b_2}\sum_n^\infty\lambda^{n+1}\sigma^n_{n+A,n+B}\frac{\Lambda^{n+B}}{w^{n+A}}\phi^4\qquad (B<A\leq1)\label{e36}\ee
(with the sum over $n$ such that $n+A\geq1$, $n+B>0$, $n\geq1$), then we see that portions of $S_n(\xi)$ are being determined beyond $n=4$, even though we only have the complete expression for $S_0(\xi)$ to $S_4(\xi)$ when only the five loop contributions to the RG functions are at our disposal. This is because $S_0(\xi)$ to $S_{p-1}(\xi)$ do not constitute the solution to the RG equation (eq. \refer{e4}) when the RG functions are known to $p$ loop order.

We do not provide the closed form expression for all of the $V_{A,B}$ as they are too long. In appendix \refer{appa} though, the sums required are worked out. Some of the simpler contributions to $V$ are
\bea V_{1,0}&=&\frac{1}{b_2}\sum_{n=0}^\infty\lambda^{n+1}\lb\rho^n\sigma^0_{1,0}\rb\frac{\Lambda^n}{w^{n+1}}\phi^4\quad\mbox{ by eqs. (\ref{e18},~\ref{e36})}\nonumber\\
&=&\frac{\lambda}{b_2}\frac{\sigma^0_{1,0}\phi^4}{w-\lambda\rho\Lambda}\quad\mbox{ by eq. \refer{eA1}}\label{e37}\eea
and
\bea V_{0,-1}&=&\frac{1}{b_2}\sum_{n=1}^\infty\lambda^{n+1}\lb-\frac{4g_2}{b_2}\rho^{n-1}\sigma^0_{1,0}\rb\frac{\Lambda^{n-1}}{w^n}\phi^4\nonumber\\
&=&\frac{-4g_2\lambda^2}{b_2^2}\frac{\sigma^0_{1,0}\phi^4}{w-\lambda\rho\Lambda}\quad\mbox{ by eq. \refer{eA1}}.\label{e38}\eea
Eqs. (\ref{e37},~\ref{e38}) serve to demonstrate that the singularity in \mcv is shifted away from the ``Landau singularity'' $w=0$ (as is implied by eq. \refer{e12}) to \be w-\lambda\rho\Lambda=\lsb1-\frac{b_2}{2}\lambda\ln\frac{\phi^2}{\mu^2}\rsb+
\lambda\frac{b_3}{b_2}\ln\left|1-\frac{b_2}{2}\lambda\ln\frac{\phi^2}{\mu^2}\right|=0.\ee
In Ref. \cite{c10}, $S_0,~ \cdots,~ S_4$ have been used to estimate the Higgs mass in the conformal limit of the Standard Model.  The contributions to $S_5$ and beyond considered here
may further refine these estimates.

\section{An Alternate Summation}

We now write \mcv in the form \cite{c8,c13}
\be V=Y(\lambda,l)\phi^4\qquad (l=\frac{1}{2}L=\ln\frac{\phi}{\mu})\label{e40}\ee
and in place of eq. \refer{e6} we group terms in the expansion of eq. \refer{e2} in the form
\be Y(\lambda,l)=\sum_{n=0}^\infty A_n(\lambda)l^n \label{e41}\ee
so that $A_n(\lambda) = \displaystyle{\sum_{m=n}^\infty} \lambda^{m+1}2^{-n}T_{m,n}$, a sum which contains contributions coming from all orders in the loop expansion.
Substitution of eq. \refer{e41} into eq. \refer{e4} leads to the recursion relation
\be (n+1)A_{n+1}(\lambda)=\lb\hat{\beta}\frac{\dd}{\dd\lambda}+4\hat{\gamma}\rb A_n(\lambda)\label{e42}\ee
by considering the coefficients of terms of order $l^n$ to vanish. We have defined $\hat{\beta}=\beta/(1-\gamma)$, $\hat{\gamma}=\gamma/(1-\gamma)$ in eq. \refer{e42}. If now we set
\be \hat{A}_n(\lambda)=A_n(\lambda)\exp\lb4\int^\lambda_{\lambda_0}\dd x \frac{\hat{\gamma}(x)}{\hat{\beta}(x)}\rb\ee
and define
\be \eta(\lambda)=\int^\lambda_{\lambda_0}\frac{\dd x}{\hat{\beta}(x)}=\int^\lambda_{\lambda_0}\frac{1-\gamma(x)}{\beta(x)}\dd x\label{e46v2} \ee
and let $\lambda(\eta)$ to be the inverse function of $\eta(\lambda)$, we find that eq. \refer{e42} becomes
\be (n+1)\hat{A}_{n+1}(\eta)=\frac{\dd}{\dd\eta}\hat{A}_n(\eta),\label{eq47}\ee
where the dependence of $\hat{A}_m(\eta)$ on $\eta$ is realized through $\lambda(\eta)$, i.e., $\hat{A}_m(\eta)=\hat{A}_m(\lambda(\eta))$.
Eq. \refer{eq47} can be iterated to give
\be \hat{A}_n(\eta)=\frac{1}{n!}\frac{\dd^n}{\dd\eta^n}\hat{A}_0(\eta). \label{e46}\ee
By eq. \refer{e46}, the expansion of eq. \refer{e4} becomes
\bea Y(\lambda,l)&=&\sum_{n=0}^\infty\frac{l^n}{n!}\lsb \frac{\dd^n}{\dd\eta^n}\hat{A}_0(\lambda(\eta))\rsb \exp\lb-4\int^{\lambda(\eta)}_{\lambda_0}\frac{\hat{\gamma}(x)}{\hat{\beta}(x)}
\dd x\rb\nonumber\\
&=&\hat{A}_0(\lambda(\eta+l))\exp\lb-4\int^{\lambda(\eta)}_{\lambda_0}\frac{\hat{\gamma}(x)}{\hat{\beta}(x)}\dd x\rb\nonumber\\
&=&A_0(\lambda(\eta+l))\exp\lb4\int^{\lambda(\eta+l)}_{\lambda(\eta)}\frac{\gamma(x)}{\beta(x)}\dd x\rb.\label{e47}\eea
Consequently, $V$ is determined by $A_0(\lambda)$ which is the sum of all contributions to $V$ that are independent of $\ell$.  (One goes to arbitrary high order in the loop expansion in
this sum.)  This is in keeping with the result of section II where it is shown that $S_n(\xi)$ can be determined by the RG equation once the values of $T_{0,0},~ \cdots,~ T_{n,0}$ are known,
as $A_0(\lambda)$ is determined by $T_{n,0} ~(n = 0, ~1,~ \cdots)$.  To find $A_0(\lambda)$ we need an extra condition on \mcv; we turn to the requirement that \mcv have a minimum at the vacuum
expectations value $v$ of $\phi$.
If now $\mu$, the mass scale, is taken to be $v$, the vacuum expectation value of \mphi (viz  at $\phi=v$, \mcv is minimized), then we have
\be\left. \frac{\dd V}{\dd \phi}\right|_{\phi=v}=0. \label{eq50}\ee
Since $l=0$ when $\mu=\phi$, this condition and the expansion of eq. \refer{e41} lead to
\be \lsb A_1(\lambda)+4A_0(\lambda)\rsb v^3=0,\label{e49}\ee
so that if $v\neq0$, $A_1(\lambda)=-4A_0(\lambda)$. Strictly speaking, this has only been shown at the value of $\lambda$ corresponding to $\mu=v$ but as this value is not fixed, we have a functional relation between $A_1$ and $A_0$; eq. \refer{e49} must hold irrespective of the value of $\lambda$ at this particular value of $\mu$.  (In Ref. \cite{c1}, a similar equation, eq. (4.8), was used to relate the
four point coupling to the gauge coupling in scalar QED when at one-loop order. Here, however, eq. \refer{e49} is used to relate the functions $A_1(\lambda)$ and
$A_0(\lambda)$, not to fix the value of $\lambda$ at $\mu = v$.) This relation, when combined with eq. \refer{e42} when $n=0$, gives
\be \lsb \hat{\beta}\frac{\dd}{\dd\lambda}+4(1+\hat{\gamma})\rsb A_0=0\label{eq52}\ee whose solution is
\be A_0(\lambda)=A_0(\lambda_0)\exp\lb-4\int_{\lambda_0}^\lambda\frac{\dd x}{\beta(x)}\rb. \label{eq53}\ee
Eq. \refer{e47} then becomes
\be Y(\lambda,l)=A_0(\lambda_0)\exp\lb -4\int_{\lambda_0}^{\lambda(\eta)}\frac{\dd x}{\beta(x)}\rb\exp\lb-4\int_{\lambda(\eta)}^{\lambda(\eta+l)}\frac{1-\gamma(x)}{\beta(x)}\dd x\rb \ee
which by eq. \refer{e46v2} becomes
\bea Y(\lambda,l)&=&A_0(\lambda_0)\exp\lb -4\int_{\lambda_0}^\lambda\frac{\dd x}{\beta(x)}\rb \exp\lsb-4\lb(\eta+l)-\eta\rb\rsb\nonumber\\
&=&A_0(\lambda_0)\exp\lb -4\int_{\lambda_0}^\lambda\frac{\dd x}{\beta(x)}\rb\lb\frac{\mu}{\phi}\rb^4.\label{e53}\eea
Substitution of eq. \refer{e53} into eq. \refer{e40} shows that all dependence of \mcv on \mphi cancels, provided $v\neq0$ in eq. \refer{e49}. The dependence of \mcv on \mmu and \mlambda resulting from eq. \refer{e53} ensures that eq. \refer{e4} is satisfied.

We also note that the condition of eq. \refer{eq50}, if we do not choose $\mu$ to be equal to $v$, leads to
\be \sum_{n=0}^\infty A_n (\lambda) \left[4\ln^n\left(\frac{v}{\mu}\right) + n \ln^{n-1} \left(\frac{v}{\mu}\right)\right]v^3 = 0.\ee
If $v \neq 0$, and if this were to hold at each order in $\ln \left(\frac{v}{\mu}\right)$ (as is the case with $\mu$ and $\lambda(\mu,\lambda_0)$ being independent variables) then
\be A_{n+1} (\lambda) = \frac{-4}{n+1} A_n(\lambda)\ee
and so
\be A_n (\lambda) = \frac{(-4)^n}{n!} A_0(\lambda).\ee
The sum in eq. \refer{e41} again becomes
\be Y(\lambda ,\ell) = \sum_{n=0}^\infty \frac{A_0(\lambda)}{n!} (-4\ell)^n,\ee
reproducing the result of eq. \refer{e53} once eq. \refer{eq53} is taken into account.

We can also consider the consequence of substituting the expansion of eq. \refer{e41} into the CW renormalization condition of eq. \refer{e3}.  We then obtain
\be 24A_4(\lambda) + 60A_3(\lambda) + 70A_2(\lambda) + 50A_1(\lambda) + 24A_0(\lambda) = 24\lambda. \label{eq60}\ee
Upon iterating the recursion relation of eq. \refer{e42}, we find that the RG equation implies that
\be A_{n+1} (\lambda) = \frac{1}{n!} \left(\hat{\beta} \frac{\dd}{\dd\lambda} + 4\hat{\gamma}\right)^n A_0(\lambda);\label{eq61} \ee
substitution of eq. \refer{eq61} into eq. \refer{eq60} results in
\be \left(\hat{\beta} \frac{\dd}{\dd\lambda} + 4\hat{\gamma} + 4\right) \left(\hat{\beta} \frac{\dd}{\dd\lambda} + 4\hat{\gamma} + 3\right)
\left(\hat{\beta} \frac{\dd}{\dd\lambda} + 4\hat{\gamma} + 2\right) \left(\hat{\beta} \frac{\dd}{\dd\lambda} + 4\hat{\gamma} + 1\right)
A_0(\lambda) = 24\lambda .\label{eq62}\ee
Since
\be \frac{1}{(x+4)(x+3)(x+2)(x+1)} = -\frac{1}{6} \frac{1}{x+4} + \frac{1}{2} \frac{1}{x+3} - \frac{1}{2} \frac{1}{x+2} + \frac{1}{6} \frac{1}{x+1}\label{eq63}
\ee
and because the solution to $\left(\hat{\beta} \frac{d}{d\lambda} + 4\hat{\gamma} + a\right)f_a(\lambda) = g(\lambda)$ is
\be f_a(\lambda) = \exp\left(- \int_{\lambda_0}^\lambda \dd x \frac{4\hat{\gamma}(x) + a}{\hat{\beta}(x)}\right)
\left[ C_a + \int_{\lambda_0}^\lambda \dd x \frac{g(x)}{\hat{\beta}(x)} \exp \left( \int_{\lambda_0}^x dy \frac{4\hat{\gamma}(y)+a}{\hat{\beta}(y)}\right)\right]
\ee
we see that eq. \refer{eq62} has the solution
\bea A_0(\lambda)& =& \exp\left(-4 \int_{\lambda_0}^\lambda \dd x \frac{\hat{\gamma}(x)}{\hat{\beta}(x)}\right)
\sum_{a=1}^4 K_a \exp \left(-a \int_{\lambda_0}^\lambda \frac{\dd x}{\hat{\beta}(x)}\right)\nonumber\\
&&\times\left[ C_a + 24 \int_{\lambda_0}^\lambda \dd x \frac{x}{\hat{\beta}(x)} \exp \left(
\int_{\lambda_0}^x dy \frac{4\hat{\gamma}(y)+a}{\hat{\beta}(y)}\right)\right] \eea
where, by eq. \refer{eq63} $K_4 = -K_1 = \frac{1}{6}$, $K_3 = -K_2 = \frac{-1}{2}$ and the constants $C_a~ (a = 1,~\cdots,~ 4)$ are not fixed.

We do see from eq. \refer{eq52} that the solution given by eq. \refer{eq53} is consistent with the CW renormalization condition as expressed in eq. \refer{eq62} provided $24\lambda = 0$; that this
the coupling vanishes and the theory is ``trivial'' .  Triviality has also been discussed for the massless $\lambda\phi^4$ model in other contexts \cite{c15}.  We note that
having a non-vanishing vacuum expectation value $v$ for $\phi$ is not precluded by having a trivial theory, though $v$ is no longer determined by minimizing $V$.

The results of section II can be seen to support the result that $V$ is in fact independent of $\phi$.  We adopt the approach of Ref. \cite{c10}, setting
\be V_p = \sum_{n=0}^p \lambda^{n+1} S_n(\lambda L)\phi^4 + \pi^2 K_p \phi^4\qquad~ (p = 0,~1,~2,~\cdots)\label{eq66} \ee
where $V_p$ is the sum of the N$^p$LL contribution to $V$ and a ``counter-term'' $\pi^2 K_p = \displaystyle{\sum_{n=p+1}^\infty} \lambda^{n+1} T_{n,0}$ contains all log-independent
contributions to $V$ coming from $(p+1)$-loop order and beyond.  If we scale the mass parameter $\mu$ so that $\mu = v = 1$ (with $v$ being the vacuum expectation value of
$\phi$) then there are two undetermined parameters in $V_p$, namely $K_p$ and $\lambda$.  We first express $K_p$ in terms of $\lambda$ by eq. \refer{e3} and then fix $\lambda$ by
eq. \refer{eq50}.  ($S_0$ and $S_1$ are given by eqs. (9, 10) and $S_2$, $S_3$, $S_4$ appear in Ref. \cite{c10}.)  We only accept solutions in which $\lambda$ is non-negative as being
physical.  In Table \refer{tb1} we provide the values of $\lambda$, $K_p$ coming from $V_p~ (p = 0,~ \cdots,~ 4)$ as well as the value of $V_p$ when $\phi = v$ and the value of
$(\frac{\phi}{v})$ when $V_p$ becomes singular; both for the N=1 and N=4 O(N) versions of the massless $\lambda\phi^4$ model.
\begin{table}
\begin{tabular}{|c|c|c|c|c|c|c|c|c|} \hline
 $p$ &\multicolumn{2}{c}{$\lambda$}&\multicolumn{2}{|c}{$K_p$} &\multicolumn{2}{|c} {$\displaystyle  \min V_p(1)/v^4$} & \multicolumn{2}{|c|}{$\displaystyle \frac{\phi}{v}$ at singularity}\\  \hline\hline
 & N=1 & N=4 & N=1 & N=4 & N=1 & N=4 & N=1 & N=4
\\ \hline
0 &0.712 &0.534 & -0.0780 & -0.0585 & -0.0578 & -0.0434 & 21.7 & 21.7 \\ \hline
1 &  0   &   0  &     0  &    0   &   --  &   --  &  --  &  --  \\ \hline
2 &0.545 &0.417 & -0.0514 & -0.0390 & -0.0387 & -0.0296 & 51.0 & 51.8 \\ \hline
3 &  0   &   0  &     0  &    0   &   --  &   --  &  --  &  --  \\ \hline
4 &0.458 &0.354 & -0.0420 & -0.0321 & -0.0296 & -0.0228 & 120 & 105 \\ \hline
\end{tabular}
\caption{Coupling constant, counter term and potential minimum and singularity at different orders in the CW Scheme. \label{tb1} }
\end{table}

\begin{figure}
\includegraphics[scale=0.4]{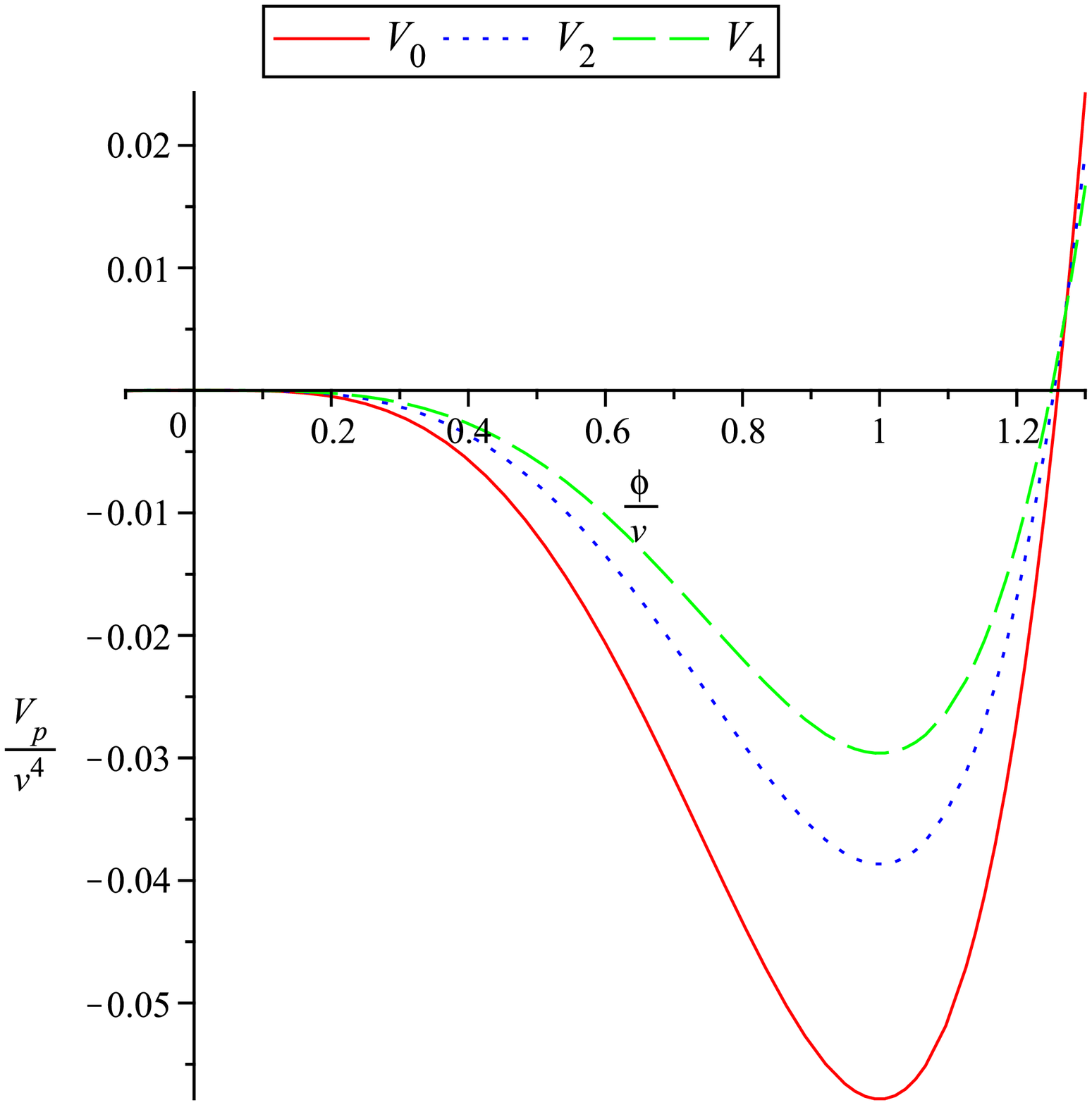}
\includegraphics[scale=0.4]{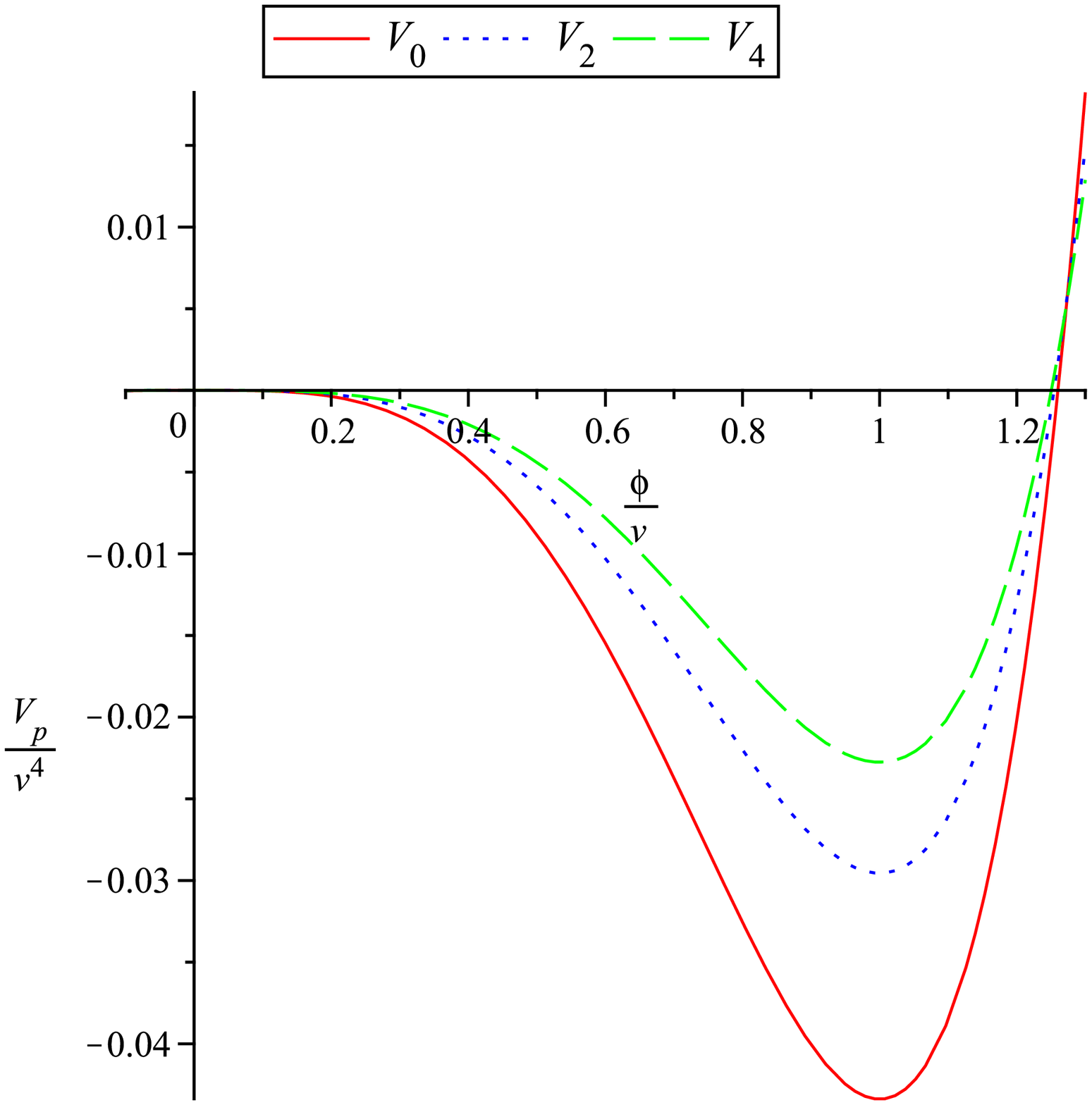}
\caption{The effective potential for N=1 (left) and N=4 (right) O(N) $\lambda\phi^4$ theory at different order $p$ in the CW scheme. \label{fig1}}
\end{figure}

In Fig. \refer{fig1} we plot $V_p$ for $p = 0, 2, 4$ for N=1 and N=4 respectively in the region between $\phi=0$ and $\phi\sim v$.  (Fig. \refer{fig1} (right) appears in Ref. \cite{c10}.)  It is apparent that when $V_p$ is computed in the CW scheme,
having $p = 1, 3$ leads to having $V_p$ flat in this region if it is to be physical--i.e. there are no non-negative values of $\lambda$ other than zero that are physically acceptable.  If however,
$p = 0, 2, 4$ there are physically acceptable positive values for $\lambda$, but as $p$ increases, these values decrease, resulting in the potential becoming flatter and the singularities receding away from $\phi = 0$. If one consider the region between $\phi=v$ and the singularities for $p=0,~2,~4$, $V_p(\phi)$ increases at the same rate for all $p$ until the region about their respective singularities. This we take to be a strong indication that as $p$ increases, $V_p$ approaches a flat potential
with vanishing coupling--the ``trivial'' theory considered above.

The argument for the effective potential in the massless $\lambda\phi^4$ model being flat given in Ref. \cite{c8} was given in the context of the $\overline{\mbox{MS}}$ renormalization scheme.
So far, here we have employed the CW renormalization scheme.  We now will see if the $\overline{\mbox{MS}}$ analogues of the $V_p$ given in eq. \refer{eq66} support the results of Ref. \cite{c8} as $p$ increases.

The form of the potential when $\overline{\mbox{MS}}$ is used to compute $V$ in this model is
\be V = \sum_{n=0}^\infty \sum_{m=0}^n \lambda^{n+1} \overline{T}_{n,m}\overline{L}^m \phi^4\label{eq67}\ee
where now $\overline{L} = \ln \left(\frac{\lambda\phi^2}{\overline{\mu}^2}\right)$ where $\overline{\mu}^2$ is the mass scale parameter occuring in the derivation of eq. \refer{e13v2}.
Regrouping the double sum of eq. \refer{eq67} in a manner analogous to eq. \refer{e6}, we have
\be\overline{S}_n(\overline{\xi}) = \sum_{m=0}^\infty \overline{T}_{n+m,m} \overline{\xi}^m\qquad (\overline{\xi} = \lambda \overline{L})\ee
and now in place of eq. \refer{e7} we have
\begin{subequations}
\begin{eqnarray}
&&(-2+\overline{b}_2\overline{\xi}) \overline{S}_0^\prime + (\overline{b}_2 + 4\overline{g}_1)\overline{S}_0 = 0\\
&&(-2+\overline{b}_2\overline{\xi})\overline{S}_1^\prime + (2\overline{b}_2 + 4\overline{g}_1)\overline{S}_1 +
(2\overline{g}_1 +  \overline{b}_2 + \overline{b}_3\overline{\xi})S_0^\prime+(\overline{b}_3 + 4\overline{g}_2) \overline{S}_0 = 0\\
&&(-2+\overline{b}_2\overline{\xi}) \overline{S}_2^\prime + (4\overline{g}_1 + 3\overline{b}_2)\overline{S}_2 +(2\overline{g}_1 +\overline{b}_2 + \overline{b}_3 \overline{\xi})\overline{S}_1^\prime\nonumber\\
&&+(2\overline{b}_3 + 4\overline{g}_2)\overline{S}_1 + (2\overline{g}_2 + \overline{b}_3 + \overline{b}_4 \overline{\xi})S_0^\prime + (4\overline{g}_3 + \overline{b}_4)S_0 = 0.
\end{eqnarray}
\end{subequations}
These can be solved in turn, with the boundary condition $\overline{S}_n(0) = \overline{T}_{n,0}$.  There is no equivalent to eq. \refer{e10} in the $\overline{\mbox{MS}}$ scheme for determining
$\overline{T}_{n,0}$; however the two-loop calculation of $V$ using $\overline{\mbox{MS}}$ appearing in Ref. \cite{c11} allows one to read off $\overline{T}_{0,0}$,
$\overline{T}_{1,0}$, $\overline{T}_{2,0}$.  (A three-loop calculation of $V$ for N=1 using an on-shell renormalization scheme appears in Ref. \cite{c16}.)  We see that
\begin{subequations}
\begin{eqnarray}
\overline{T}_{0,0} &=& 1\\
\overline{T}_{1,0} &=& \frac{1}{(4\pi)^2}\frac{1}{4} \left[ (12)^2 (\ln 12 - 3/2) +  (N-1)(4)^2(\ln 4 - \frac{3}{2})\right]\\
 \overline{T}_{2,0} &=& \frac{1}{(4\pi)^4} \left\lbrace \frac{1}{8} (24)^2 (12)(5 + 8\Omega(1) - 4\ln 12 + \ln^212)\right. \nonumber \\
&+& \frac{1}{72} (N-1)(24)^2(12+2(4))\left[5+8\Omega(3) - 4\ln 4 + \ln^2 4\right.\nonumber \\
&+&\left. \frac{2(12)}{(12)+2(4)} (\ln 3)(\ln 4 - 2)\right] \nonumber \\
&+& [3(12)^2(1-\ln 12)]^2 + (N^2-1)(4)^2(1-\ln 4)^2\nonumber\\
&+&\left.  2(N-1)(12)(4) [1-\ln 4 - \ln 12 + \ln 4 \ln 12]\right\rbrace
\end{eqnarray}
\end{subequations}
(where $\Omega(\Delta) = \frac{\sqrt{\Delta(4-\Delta)}}{\Delta + 2} \int_0^\theta \ln(2\sin x)\dd x$, $\sin\theta = \frac{\sqrt{\Delta}}{2}$ for $\Delta < 4$).  With these numerical
values for $\overline{T}_{0,0}$, $\overline{T}_{1,0}$, $\overline{T}_{2,0}$ we can solve for $\overline{S}_0$, $\overline{S}_1$, $\overline{S}_2$ and form
\be \overline{V}_p = \sum_{n=0}^p \lambda^{n+1} \overline{S}_n(\lambda\overline{L})\phi^4 + \pi^2 \overline{K}_p\phi^4 \label{eq71}\ee
in analogy with eq. \refer{eq66}.

The coupling $\lambda$ in the $\overline{\mbox{MS}}$ is not to be identified with the ``physical'' coupling any more than a mass parameter is to be identified with a pole mass in this
renormalization scheme.  For $\lambda$ to be the physical coupling, we again employ eq. \refer{e3}.  This condition and eq. \refer{eq50} again can be used to fix $\lambda$ and $\overline{K}_p$ in
eq. \refer{eq71} as in the CW scheme if $\lambda$ is to be a ``physical'' coupling.

We find that in contrast with the CW scheme, when $p = 1$ there exist a positive solution for $\lambda$.  This indicates a renormalization scheme dependence in our perturbative
analysis.  However, in Table \refer{tb2} where we have presented the $\overline{\mbox{MS}}$ results corresponding to the CW results appearing in Table \refer{tb1}, it is again apparent that as $p$ increases,
$\overline{V}_p$ becomes increasingly flat in the region between the two singularities and the singularities again recede away from $\phi = 0$.

\begin{table}
\begin{tabular}{|c|c|c|c|c|c|c|c|c|} \hline
 $p$ &\multicolumn{2}{c}{$\lambda$}&\multicolumn{2}{|c}{$K_p$} &\multicolumn{2}{|c}{$\displaystyle \min\frac{V_1(1)}{v^4}$} & \multicolumn{2}{|c|}{$\displaystyle \frac{\phi}{v}$ at singularity}\\ \hline\hline
 & N=1 & N=4 & N=1 & N=4 & N=1 & N=4 & N=1 & N=4
\\ \hline
0 &0.768 &0.617 & -0.0806 & -0.0622 & -0.0614 & -0.0486 & 19.9 & 18.3 \\ \hline
1 &0.669 &0.546 & -0.0787 & -0.0603 & -0.0561 & -0.0446 & 32.5 & 27.5 \\ \hline
2 &0.645 &0.528 & -0.0751 & -0.0578 & -0.0533 & -0.0426 & 37.4 & 31.0 \\ \hline
\end{tabular}
\caption{Coupling constant, counter term and potential minimum and singularity at different orders in the $\overline{\mbox{MS}}$ Scheme.\label{tb2}}
\end{table}
\begin{figure}
\includegraphics[scale=0.4]{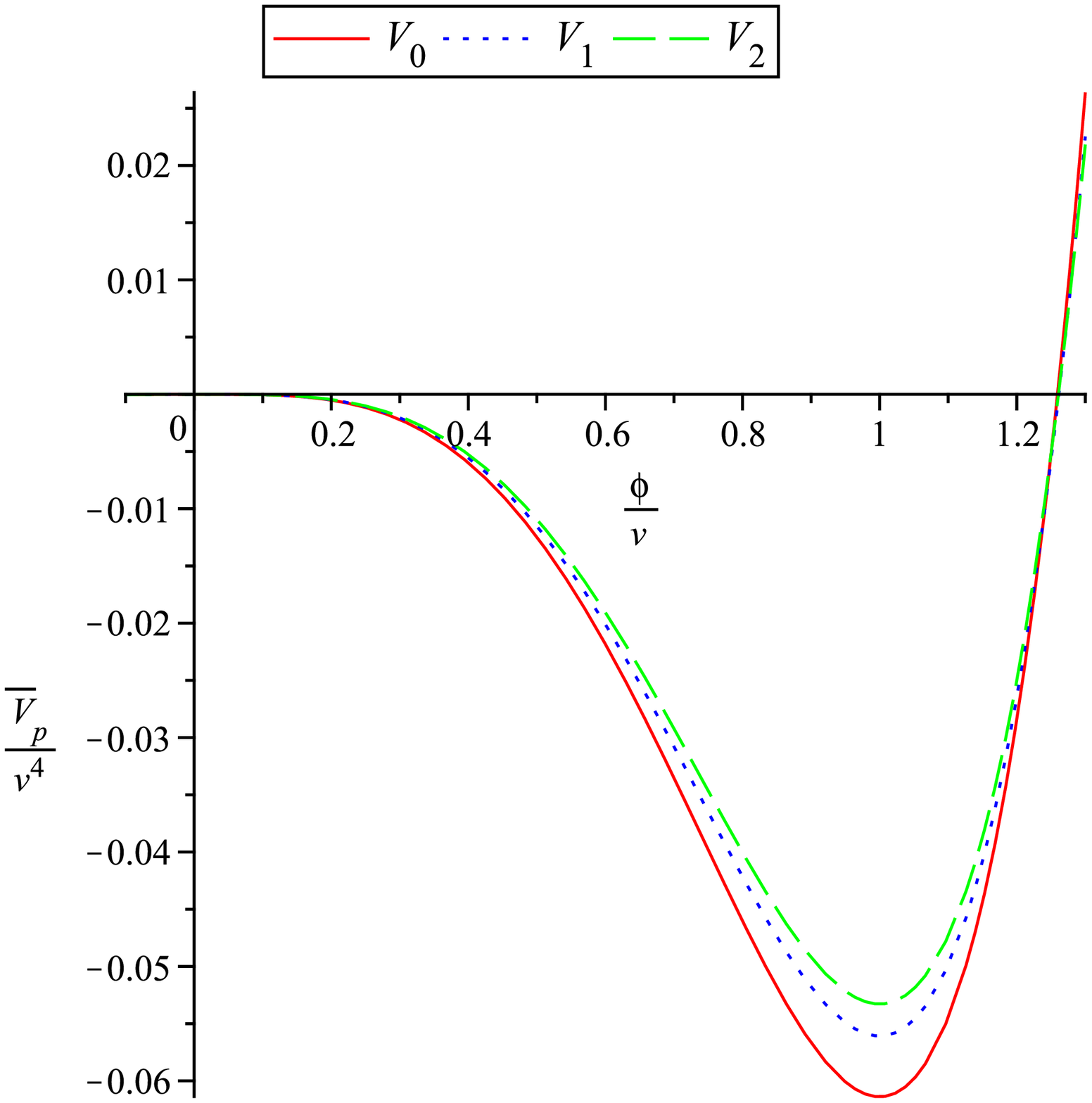}
\includegraphics[scale=0.4]{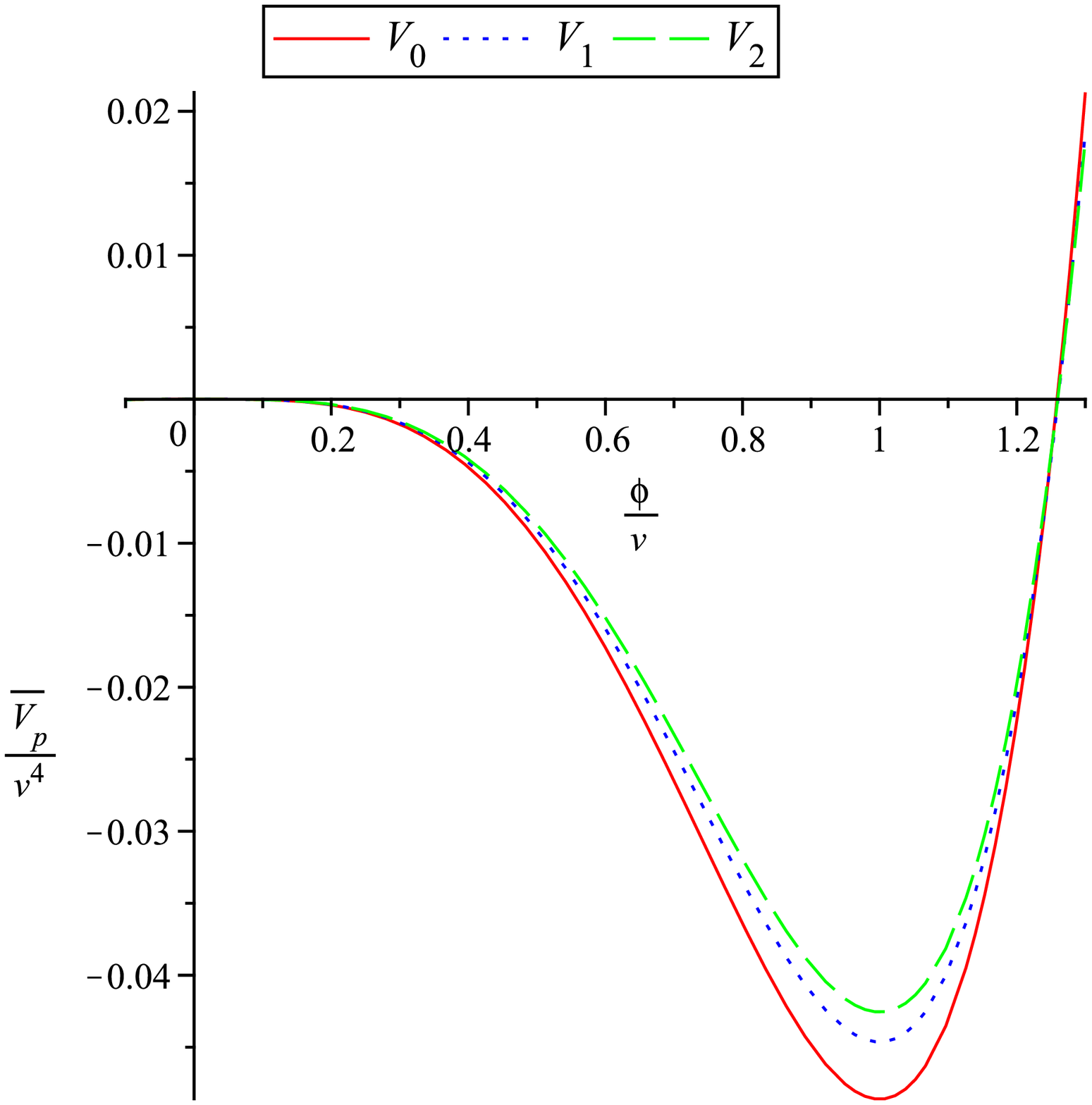}
\caption{The effective potential for N=1 (left) and N=4 (right) O(N) $\lambda\phi^4$ theory at different order $p$ in the $\overline{\mbox{MS}}$ scheme. \label{fig2}}
\end{figure}

The plot of $\overline{V}_p~(p =0,1,2)$ in Fig. \refer{fig2} for N=1
and N=4 respectively again can be taken as an indication that as $p$ increases, the potential approaches being flat, consistent  with Ref. \cite{c8}.

\section{Discussion}

The renormalization group provides a way of computing parts of radiative corrections to physical quantities beyond the order in the loop expansion to which explicit calculations have been performed. We have shown how a one loop calculation of the RG functions, when using the CW scheme, can be used to fix the sum of all leading-log contribution to the effective potential \mcv (with these contributions coming from all orders in the loop expansion). This procedure can then be used to sum all N$^{\mbox{\scriptsize p}}$LL contributions to \mcv using the $p+1$ order expression for the RG functions. Furthermore, we have demonstrated how the RG equation, when the RG functions are known to order $p$, can be used to fix at least portions of the effective potential coming from terms of order N$^{\mbox{\scriptsize p+A-1}}$LL in the expansion of \mcv $(A=1,~2,~3,~\cdots)$. (This is because the solution to the RG equation when the RG functions are known to order $p$ is not merely given by $S_0$ to $S_{p-1}$.) We have found that in these sums, the singularity in $V$ is shifted away from the usual ``Landau'' singularity.

In addition, we have shown how the RG equation can be used to fix \mcv in terms of the log-independent portion of \mcv, and that when this is combined with the condition that \mcv be minimized when $\phi=\mu$, then all dependence of \mcv on $\phi$ disappears if there is spontaneous symmetry breaking.  We also see that in the CW renormalization scheme, the coupling vanishes and the
theory becomes ``trivial''.  This is supported by plots of the contributions from the $LL$ etc. contributions to $V$ in the CW and $\overline{\mbox{MS}}$ renormalization schemes.

It would be interesting to generalize this procedure to models in which there is a classical mass for $\phi$, or more than one coupling constant so that this technique could be applied to computing \mcv in the Standard Model beyond the order which has been considered in Ref. \cite{c10}. It would also be useful to examine radiative corrections to other physical processes using the techniques developed here \cite{c14}.

\begin{acknowledgments}
We would like to thank R.B. Mann and T.G. Steele for helpful suggestions. One of the authors (J. J) was supported by the Natural Sciences and Engineering Research Council of Canada. Roger Macleod raised a useful point.
\end{acknowledgments}

\appendix
\section{Appendix A}\label{appa}
The following sums are needed to compute $V_{A,B}$ in eq. \refer{e36}; they converge provided $|x|<1$.
\begin{enumerate}
\item \be \sum_{n=a}^\infty x^n=\frac{x^a}{1-x} ;\label{eA1}\ee
\item \be \sum_{n=a}^\infty n x^n=x\frac{\dd}{\dd x}\frac{x^a}{1-x}=\frac{ax^a+(1-a)x^{a+1}}{(1-x)^2}; \label{eA2}\ee
\item \be \sum_{n=a}^\infty n^2 x^n=\lb x^2\frac{\dd^2}{\dd x^2}+x\frac{\dd}{\dd x}\rb\frac{x^a}{1-x}=\frac{a^2x^a+(1+2a-2a^2)x^{a+1}+(1-a)^2x^{a+2}}{(1-x)^3}; \ee
\item \bea \sum_{n=2}^\infty n\lb\frac{1}{2}+\frac{1}{3}+\cdots+\frac{1}{n}\rb x^n &=& x\frac{\dd}{\dd x}\sum_{n=2}^\infty \int_0^1\dd t(t+t^2+\cdots+t^{n-1})x^n\nonumber\\
    &=&x\frac{\dd}{\dd x}\sum_{n=2}^\infty \int_0^1\dd t\sum_{n=2}^\infty\lb\frac{t-t^n}{1-t}\rb x^n\nonumber\\
    &=& x\frac{\dd}{\dd x} \int_0^1\frac{\dd t}{t}\lsb (1-t)\frac{x^2}{1-x}-\frac{(x(1-t))^2}{1-x(1-t)}\rsb\nonumber\\
    &=& -\frac{x\ln(1-x)}{(1-x)^2};\label{eA4}\eea
\item \bea && \sum_{n=3}^\infty n(n-1)\lb\frac{1}{3}+\cdots+\frac{1}{n}\rb x^n \nonumber\\
&=&-\frac{1}{2}\sum_{n=3}^\infty(n^2x^n-nx^n)+x^2\frac{\dd}{\dd x}\lsb \frac{1}{x}\lb-x^2+\sum_{n=2}^{\infty} n \lb\frac{1}{2}+\frac{1}{3}+\cdots+\frac{1}{n}\rb x^n\rb\rsb\nonumber\\
&&\mbox{which by eqs. \refer{eA2}-\refer{eA4} becomes}\nonumber\\
&=&-\frac{1}{2}\lsb \frac{9x^3-11x^4+4x^5}{(1-x)^3}-\frac{3x^3-2x^4}{(1-x)^2}\rsb-x^2\frac{\dd}{\dd x}\lsb x+\frac{\ln(1-x)}{(1-x)^2}\rsb\nonumber\\
&=&\frac{-2x^2\ln(1-x)}{(1-x)^3};\eea
\item \bea && \sum_{n=2}^\infty (n+1)n\lsb\frac{1}{3}\cdot\lb\frac{1}{2}\rb+\frac{1}{4}\cdot\lb\frac{1}{2}+\frac{1}{3}\rb+\cdots+ \frac{1}{n+1}\lb\frac{1}{2}+\frac{1}{3}+\cdots+\frac{1}{n}\rb\rsb x^{n+1}\nonumber\\
&=&x\frac{\dd^2}{\dd x^2}\sum_{n=2}^\infty \int_0^1\dd\tau\lsb \frac{1}{3}(\tau)+\frac{1}{4}(\tau+\tau^2)+\cdots+\frac{1}{n+1}(\tau+\tau^2+\cdots+\tau^{n-1})\rsb x^{n+1}\nonumber\\
&=&x\frac{\dd^2}{\dd x^2} \int_0^1\dd\tau \sum_{n=2}^\infty\lsb\frac{1}{3}\frac{\tau-\tau^2}{1-\tau}+\frac{1}{4}\frac{\tau-\tau^3}{1-\tau}
+\cdots+\frac{1}{n+1}\frac{\tau-\tau^n}{1-\tau}\rsb x^{n+1}\nonumber\\
&=&x\frac{\dd^2}{\dd x^2} \int_0^1\frac{\dd \tau}{1-\tau}\sum_{n=2}^\infty\lsb\tau\int_0^1\dd t(t^2+t^3+\cdots+t^n)-\frac{1}{\tau} \int_0^{\tau}\dd t(t^2+t^2+\cdots+t^n)\rsb x^{n+1}\nonumber\\
&=&x\frac{\dd^2}{\dd x^2}\lsb x\int_0^1\frac{\dd\tau}{1-\tau}\sum_{n=2}^\infty\lb\tau\int_0^1\dd t\frac{t^2-t^{n+1}}{1-t}-\frac{1}{\tau} \int_0^{\tau}\dd t\frac{t^2-t^{n+1}}{1-t}\rb x^n\rsb\nonumber\\
&=&x\frac{\dd^2}{\dd x^2}\lsb x\int_0^1\frac{\dd\tau}{1-\tau}\lb\tau\int_0^1\frac{\dd t}{1-t}\lb\frac{t^2x^2}{1-t}-\frac{t(xt)^2}{1-xt}\rb -\frac{1}{\tau}\int_0^\tau \frac{\dd t}{1-t}\lb\frac{t^2x^2}{1-t}-\frac{t(xt)^2}{1-xt}\rb \rb\rsb\nonumber\\
&=&-\frac{x}{3(1-x)^3}\lcb 3\ln(1-x)\lsb1+2\ln \frac{x}{1-x}\rsb+6 \int^{\frac{x}{1-x}}_0\frac{\ln t}{1+t}\dd t +6\int_0^{1-x} \frac{\ln t}{1-t}\dd t+\pi^2\rcb. \eea
\end{enumerate}

\end{document}